\documentclass[11pt]{article}
\newcommand{\sss}{\vspace{.2in}}

\newcommand{\ad}{a^{\dagger}}
\newcommand{\adm}{a^{\dagger m}}

\newcommand\com[2]{\left\lbrack\,{#1}\, ,\,{#2}\,\right\rbrack} 
\def\br{\begin{eqnarray}}
\def\er{\end{eqnarray}}
\def\be{\begin{equation}}
\def\ee{\end{equation}}
\def\dx{\frac{d}{dx}}
\def\dy{\frac{d}{dy}}
\def\lb{\left(}
\def\rb{\right)}
\def\Blb{\left[}
\def\Brb{\right]}

\def\>{\rangle}              
\def\<{\langle}              
\def\({\left(}
\def\){\right)}

\def\hf{\frac{1}{2}}

\begin{document}
~\hfill{\footnotesize UICHEP-TH/98-8,~~\today}
\sss
\begin{center}
{\bf \Large {\Large  Coordinate Realizations of Deformed
Lie Algebras with Three Generators}}
\end{center}
\vspace{.5in}
\begin{center}
{\bf \large{
   \mbox{R. Dutt}$^{a,}$\footnote{rdutt@vbharat.ernet.in},
   \mbox{A. Gangopadhyaya}$^{b,}$\footnote{agangop@luc.edu, asim@uic.edu},
   \mbox{C. Rasinariu}$^{c,}$\footnote{costel@uic.edu} and
   \mbox{U. Sukhatme}$^{c,}$\footnote{sukhatme@uic.edu}
 }}
\end{center}
\vspace{.6in}
\noindent
a) \hspace*{.2in}
Department of Physics, Visva Bharati University, Santiniketan, India;
\\
b) \hspace*{.2in}
Department of Physics, Loyola University Chicago, Chicago, USA; \\
c) \hspace*{.2in}
Department of Physics, University of Illinois at Chicago, Chicago,
USA. \\

\begin{abstract}
Differential realizations in coordinate space for deformed Lie
algebras with three generators are obtained using bosonic creation and
annihilation operators satisfying Heisenberg commutation relations. The unified
treatment presented here contains as special cases all previously given
coordinate realizations of $so(2,1),so(3)$ and their deformations. Applications
to physical problems involving eigenvalue determination in nonrelativistic
quantum mechanics are discussed.
\end{abstract}
\newpage
\noindent
{\bf 1. Introduction:}
\sss

Lie groups and their associated algebras are extensively used in the analysis of
the symmetry properties of physical systems. For example, realizations of
$so(2,1)$ have been used to obtain the eigenvalues of many quantum mechanical
problems. Recent studies show that coordinate realizations of nonlinear Lie
algebras may also be interesting in determining eigenspectra of certain
physical problems in an algebraic approach \cite{Junker}.
The main purpose of this paper is to set up a unified approach for obtaining differential realizations in one and
two-dimensional coordinate space for nonlinear Lie algebras with
three generators.  

The deformed Lie algebras which we consider are described by
\be
[J_3,J_+]=J_+,~~~~~[J_3,J_-]=-J_-,
~~~~~[J_+,J_-]= f(J_3) ~~.
\label{1}
\ee
$J_\pm \equiv J_1\pm i J_2$ are
the well known raising and lowering operators. $f(J_3)$ is an arbitrary analytic
 function of the operator $J_3$. Note that the special choice $f(J_3)=2J_3$
 corresponds to $so(3)$ and $f(J_3)=-2J_3$ corresponds to $so(2,1)$. In terms of
 the Cartesian generators $J_1, J_2, J_3$, the commutation relations are
\be \label{so3}
[J_1,J_2]=\frac{i}{2} f(J_3),
~~~~~[J_2,J_3]=i J_1,
~~~~~[J_3,J_1]=i J_2 ~~.
\ee
The plan of this paper is as follows. In Sec. 2, we review some simple general
 properties of Lie algebras. In Sec. 3, we describe how to obtain
 realizations of eq. (\ref{1}) in terms of bosonic creation and
 annihilation operators ($\ad$ and $a$) satisfying Heisenberg commutation
 relations $[a,\ad]=1$. Although we are using the conventional notation $a$ and
 $\ad$ for these operators, they do not necessarily have to be Hermitian
 conjugates of each other. Appendix A contains a discussion of specific
 one-dimensional realizations of the Heisenberg algebra.
In particular, it is shown that realizations involving derivatives higher than
 the first can all be reduced to first and zero order. Sec. 4 contains a
 description of one-dimensional coordinate realizations of the Lie algebra given
 in eq. (\ref{1}). We show that our unified approach reproduces all previously
 known realizations in the literature \cite{Barut71, Adams, Filho, Fradkin, Beckers}.
 Two-dimensional coordinate realizations are described in Sec 5, along with
some applications involving eigenvalue determination
 for some nonrelativistic quantum mechanical potentials.

\sss
\noindent
{\bf 2. Some Properties of the Lie Algebra:}
\sss

For completeness and to establish notation, we describe some properties of Lie
algebras. Some are well-known, but others are new.

(i) The function $f(J_3)$ characterizes the Lie algebra given in eq.
 (\ref{1}). For subsequent work, it is convenient to define the function
 $g(J_3)$ as follows:
\be \label{g}
f(J_3) \equiv g(J_3)-g(J_3-1)~.
\ee
For example, for $so(3)$, $f(J_3)=2J_3$ and one gets $g(J_3)= J_3(J_3+1)$. It is
 easy to check that the function $g(J_3)$ is not unique - any periodic function
 of unit period can be added while maintaining eq. (\ref{g}). Note that the Casimir operator for the Lie
 algebra of eq. (\ref{1})is given by
\be \label{C}
C=J_-J_+ + g(J_3)=J_+J_- + g(J_3-1)~.
\ee
This observation is useful for many physical applications. For instance, we use
 it in Sec. 5 for eigenvalue determination.

(ii) The operators $J_+$ and $J_-$ satisfy the important property
\be
T(J_3)J_+ = J_+ T(J_3+1)~~,~~T(J_3)J_- = J_- T(J_3-1)
\ee
for any analytic function $T(J_3)$. This property is extensively used in
 obtaining realizations.

(iii) If operators $J_+, J_-, J_3$  satisfy the standard $so(3)$ Lie algebra, so
 do
 operators $\tilde J_+, \tilde J_-, \tilde J_3$ defined by $\tilde J_m = \sum_n
 M_{mn}J_n$ provided the matrix M satisfies $M^T M =  {1}$ and $\det M = +1$.
 Note that the elements of the matrix M do not have to be real, but if they are,
 the matrix is orthogonal. This property is very useful in relating all the
 $so(3)$ realizations currently available in the literature.

(iv) Given operators $J_1, J_2, J_3$ which satisfy the $so(3)$ Lie
 algebra, one can find operators $K_1, K_2, K_3$ which satisfy a more general
 algebra
\be
[K_1,K_2]= i q_3 K_3,
~~~~~[K_2,K_3]=i q_1 K_1,
~~~~~[K_3,K_1]=i q_2 K_2 ~~,
\ee
by choosing
$K_1 = \sqrt{q_2q_3}J_1, K_2 = \sqrt{q_3q_1}J_2, K_3 = \sqrt{q_1q_2}J_3.$ In
particular $K_1 = iJ_1,K_2 = i J_2, K_3 = J_3$ is a realization of
$so(2,1)$.

(v) Given operators $J_+, J_-, J_3$ which satisfy the standard $so(3)$ Lie
algebra, one can find operators $\tilde J_+, \tilde J_-, \tilde J_3$ which
satisfy the deformed algebra of eq. (\ref{1})\cite{Polychronakos}. These
operators are given by
\be \label{a}
\tilde J_+ = J_+A(J_3,C)~~,~~\tilde J_- = B(J_3,C)J_-~~,~~\tilde J_3 = J_3~~,
\ee
where  $C = J_-J_+ + J_3(J_3+1)$  is the Casimir operator of $so(3)$. The
form of the operators in eq. (\ref{a}) was chosen so that the two conditions
$[\tilde J_3,\tilde J_{\pm}]=\pm \tilde J_{\pm}$ are trivially satisfied.
In order to satisfy the third condition $[\tilde J_+,\tilde J_-]= f(\tilde
J_3)$, one needs functions $A(J_3,C),~B(J_3,C)$ which satisfy the following
condition:
\be \label{con}
A(J_3-1,C)B(J_3-1,C)[C -J_3(J_3-1)] - B(J_3,C)A(J_3,C)[C -(J_3+1)J_3] = f(J_3)~.
\ee
If $A(J_3,C)$ and $B(J_3,C)$ commute, this condition reduces to
\be \label{constraint}
H(J_3,C)[C -J_3(J_3+1)] = -g(J_3)+ p(J_3)~~;~ ~H(J_3,C) \equiv
 A(J_3,C)B(J_3,C)~,
\ee
where $p(J_3)$ is an arbitrary periodic function of period unity. It is
important to realize that only the product $H(J_3,C)$ is fixed by the
above constraint equation, but not the individual functions $A(J_3,C)$ and
$B(J_3,C)$. Given eq. (\ref{a}), it is sufficient to restrict our attention to
realizations of $so(3)$ in order to obtain realizations of any deformed Lie
algebra with three generators.

Note that for the special case of $so(3)$ itself, the choice $p(J_3)= C$ gives
 $H(J_3,C)=1$. The simplest choice  of factors $A(J_3,C)=B(J_3,C)=1$ reproduces
 the initial $so(3)$ realization, whereas a more general choice $B(J_3,C) =
 A^{-1}(J_3,C)$ yields a new realization. Furthermore, other choices of $p(J_3)$
 give additional new realizations of $so(3)$.  In particular, the choice
 $p(J_3)=0$ gives the realization $$\tilde J_+ =
 -J_+\frac{J_3(J_3+1)}{C-J_3(J_3+1)}~~,~~\tilde J_- = J_-~~,~~\tilde J_3 =
 J_3~~, $$ which differs from the original one only in one generator $J_+$. This
 freedom in choosing the periodic function $p(J_3)$ is analogous to gauge fixing
 in field theories.

An interesting nonlinear example using the above formalism comes from the choice
 $g(J_3) =  J_3^2 (J_3+1)^2 $ and $p(J_3)=C^2$.
This choice gives the realization
\be
\tilde J_+ = J_+[C +J_3(J_3+1)]~~,~~\tilde J_- = J_-~~,~~\tilde J_3 = J_3~~,
\ee
for the deformed Lie algebra corresponding to $f(J_3) = 4J_3^3$.

\sss
\noindent
 {\bf 3. Realizations of the Deformed Lie Algebra in
Terms of Bosonic Operators:}
\sss

In this section, we develop a procedure for obtaining realizations of the
Lie algebra defined by  eq. (\ref{1}) in terms of bosonic creation and
annihilation operators $\ad$ and $a$ which obey the Heisenberg algebra
commutator $\left[a,\ad \right]=1$. The number operator is defined by $N \equiv \ad a$.
It follows that $\left[N,\ad \right]=\ad,~\left[N,a \right]=-a$. More generally,
\be \label{adm}
\left[N,\adm \right]= m \adm ~~,~~\left[N,a^m \right]=-m a^m ~~,~~(m=0, \pm 1, \pm 2, \ldots).
\ee

To generate realizations of a deformed Lie algebra using the operators $\ad,a,
 N$,
we choose the following ansatz:
\be\label{12}
J_+=PF(N),~~~J_-=G(N)Q,~~~J_3=N + c~~,
\ee
where $c$ is a constant. $P$ and $Q$ are functions of $a$ and $\ad$ chosen to
satisfy the property
\be
\label{13}
\com{N}{P}=P,~~\com{N}{Q}=-Q~.
\ee
Clearly, from eq. (\ref{adm}) and eq. (\ref{13}), it follows that two possible
 choices for $P(a,\ad)$ are $\ad$ and $1/a$ and two possible choices for
 $Q(a,\ad)$ are $a$ and $1/\ad$. In fact, one can choose the linear combination
\be\label{14}
P=\alpha_1(N)\ad+\alpha_2(N)\frac{1}{a}~
,~~~Q=\beta_1(N)a+\beta_2(N)\frac{1}{\ad}~.
\ee
Using eq. (\ref{13}), it is easy to show that $P N^m = (N-1)^mP~,~ N^m Q = Q
(N-1)^m$, so that one has the property $PT(N)=T(N-1)P~,~T(N)Q=QT(N-1)$ for any
analytic function $T(N)$. Also, the dependence on $a$ and $\ad$ of
the products $PQ$ and $QP$ clearly comes only through the combination $\ad a =
 N$.

Our ansatz of eq. (\ref{12}) will satisfy the conditions of eq. (\ref{1})
 provided
\be \label{15}
F(N-1)G(N-1)PQ - G(N)F(N)QP = f(N+c)~~.
\ee
If $F(N)$ and $G(N)$ commute, and the above condition becomes
\be
\label{16}
H(N-1)PQ - H(N)QP = f(N+c)~,~~~H(N) \equiv F(N)G(N)~.
\ee
It only remains to determine $H(N)$ from eq. (\ref{16}). As in Sec. 2, note
 again
that the functions $F(N)$ and $G(N)$ do not appear separately but only appear as
their product $H(N)$. Also, note that in Sec. 5, we will discuss a situation
where $F(N)$ and $G(N)$ do not commute.

\sss
\noindent
{\bf 4. One-Dimensional Coordinate Realizations:}
\sss

Here we consider one-dimensional coordinate realizations for $a,\ad$ such that
$[a,\ad]=1$. Eqs. (\ref{12}), (\ref{14}) and (\ref{16}) now immediately give a
realization for the nonlinear algebra of eq. (\ref{1}). As an  example we
consider the same deformed Lie algebra with $f(J_3)=4J_3^3$ as in Section 2. We
make the simple choice $P=\ad=x, Q = a = d/dx, c = 0$ which gives $PQ = N, QP =
 N
+1, N = x d/dx$. Eq. (\ref{16}) now reads: $ H(N-1)N - H(N)(N+1) = 4 N^3$ whose
solution is $H(N)= -N^2(N+1)$. Taking $G(N) = 1$ our coordinate realization is:
\be
J_+ = -x\lb x \dx\rb^2 \lb x \dx +1 \rb ~~,~~J_- = \dx~~,~~J_3 = x \dx~~.
\ee

General coordinate realizations of $a, \ad$ are discussed in Appendix A. Any of
these can be used to generate different one-dimensional realizations of deformed
Lie algebras. Our formalism is very flexible since there is freedom in choosing
$a,\ad$ (Appendix A) and the operators $P,Q$ in eq. (\ref{14}). Furthermore,
 once
$H(N)$ has been determined from eq. (\ref{16}), one has various choices for
factorization into the functions $F(N), G(N)$ which appear in the final
realization given in eq. (\ref{12}). Our formalism contains as special cases all
the coordinate realizations published in the literature. We shall now illustrate
this statement for specific realizations discussed in \cite{Filho} and
 \cite{Barut71}.

Filho and Vaidya \cite{Filho} have discussed physical applications based on
the following representation of $so(2,1)$:
\be \label{fv}
J_+ = 2 \frac{d^2}{dy^2} - \frac{2 \alpha}{y^2}~~,
~~J_- = \frac{y^2}{8}~~,~~J_3 = -\frac{y}{2} \dy - \frac{1}{4}~~,
\ee
where $\alpha$ is an arbitrary constant. In order to obtain this realization
as a special case of our formalism, we choose $a, \ad$ by taking $\theta = 0,
 h(y) = 1/{y^2},
r(y) = -y^2/4 $ in eq. (\ref{A2}) in Appendix A. This gives $$a =
-\frac{y^3}{2} \dy - \frac{y^2}{4}~,~\ad = \frac{1}{y^2}~,~N = -\frac{y}{2}
\dy - \frac{1}{4} ~~.$$ Furthermore, choosing $P=\ad~,~Q=1/\ad$ in eq.
(\ref{14}), implies that the constraint (\ref{16}) on $H(N)$ reads
$$H(N-1)-H(N)=-2N.$$  The solution is $H(N) = N(N+1)+\beta,$  where $\beta$
is an arbitrary constant. Choosing the factorization $G(N)=1/8$ and $F(N)=8
H(N)$, eq. (\ref{12}) with $c=0$ and $\beta = (3-4\alpha)/16 $ after
 simplification gives the Filho-Vaidya
realization of eq. (\ref{fv}).

Another example of a  differential realization of the $so(2,1)$ algebra was
 given
by Barut and Bornzin \cite{Barut71}. Their expressions for the
generators are:
\be
T_1 = \frac{1}{2} \(\!\! \frac{y^{2-n}}{n^{2}} p_y^2 + \frac{\xi}{y^n} - y^n
 \!\!\),~
T_2 = \frac{1}{n} \(\!\! yp_y - i \frac{n-1}{2} \!\),~
T_3 =   \frac{1}{2} \(\!\! \frac{y^{2-n}}{n^{2}} p_y^2 + \frac{\xi}{y^n} + y^n
 \!\!\)\!\!.
\label{bb}
\ee
Here $p_y=-iy^{-1} \frac{d}{dy}y$, $n$ is an arbitrary positive integer and
 $\xi$ is an arbitrary
constant. To make contact with our formalism, using (iii) from Section 2, we
 first
rotate $T_1, T_2, T_3$ to the new operators $J_1=iT_3, J_2=-T_1 , J_3=iT_2$.
This gives
$$
J_+ = iy^n~~,~~
J_-= - i\(\frac{y^{2-n}}{n^2}\frac{d^2}{dy^2}+2\frac{y^{1-n}}{n^2} -
 \frac{\xi}{y^n} \)
~~,~~ J_3=\frac{y}{n}\frac{d}{dy} + \frac{n+1}{2n} ~~.
$$
Next, let us take $\theta=0,~h(y)=y^n$ and $r(y)=-[n(2c-1)-1]/(ny^n)~$in eq.
 (\ref{A2}) of Appendix A.
This implies
$$
a =\frac{y^{1-n}}{n}\dy -\frac{n(2c-1) - 1}{2ny^n}~,~~\ad =y^n~~,~~
N = \frac{y}{n}\dy +\frac{n+1}{2n} -c~~.
$$
Further, choosing $\alpha_1=i, \beta_2=1$ and $\alpha_2=\beta_1=0$  in eq.
 (\ref{14}), we find a
solution of eq. (\ref{16}) of the form $H(N)=b_2N^2+b_1N +b_0$ with
 $b_2=-\frac{i}{\beta_2}$,
$b_1=-i\frac{2c+1}{\beta_2}$ and $b_0=-\frac{i}{\beta_2} \Blb  (2c+1)^2/4 -\xi
 -1/(4n^2)\Brb$.
Finally, the factorization $H=FG$ with $F = 1$, concludes the proof that
eqs. (\ref{bb}) are a particular case of our formalism. Note that the initial
 rotation of generators
seems to be essential in getting the realizations of \cite{Barut71}.

Similarly, our formalism also gives the one dimensional realizations described
 in refs. \cite{Adams} and \cite{Beckers}.
\sss

\noindent
{\bf 5. Two-Dimensional Coordinate Realizations:}
\sss

In this section we will introduce realizations of   $so(2,1)$ using two
coordinates. In contrast to the one coordinate realizations, we now allow the functions $F$ and $G$ appearing in eq. (\ref{12}) to be functions of $N$ as well as an internal coordinate $x$ and its derivative $\dx$. It is important to observe that due to this generalization, the functions $F$ and $G$ no longer commute with each other, and as a result,  equation (\ref{15}) must be used.

To construct explicit realizations of $so(2,1)$, we choose
$P=\ad=\exp(i\phi)$ and $Q=\frac{1}{\ad}=\exp(-i\phi)$, i.e.
$\alpha_2=\beta_1=0$ in eq. (\ref{14}). The simplest choice of the operator $a$
 which satisfies $\com{a}{\ad}$
is $ a=-i\exp(i\phi)\frac{\partial}{\partial \phi}$. This gives $N =\ad a =
 -i\frac{\partial}{\partial \phi}$.
As a simple example, we consider
\be
\label{realization}
F(N) = \Blb - \frac{\partial}{\partial x}
    +  W\lb x, -i \frac{\partial}{\partial \phi}\rb \Brb~~,~~
G(N) = \Blb \frac{\partial}{\partial x}
    +  W\lb x, -i \frac{\partial}{\partial \phi}\rb \Brb~~,
\ee
where $W$ is a function to be determined.
Substitution in eq. (\ref{15}) yields
\be
\label{eee}
\Blb
W^2  \lb x, -i \frac{\partial}{\partial \phi}-1 \rb
- \frac{dW\lb x, -i \frac{\partial}{\partial \phi}-1\rb }{dx}  \Brb
\!- \! \Blb
W^2  \lb x, -i \frac{\partial}{\partial \phi} \rb
+ \frac{dW\lb x, -i \frac{\partial}{\partial \phi} \rb }{dx}  \Brb
= f \lb  -i \frac{\partial}{\partial \phi} +\! c \rb.
\ee

The left hand side of this equation depends on $x$ while the right hand side
does not. In order to get a two dimensional realization one needs a solution of eq.
 (\ref{eee}). In supersymmetric quantum mechanics, this equation is well known to be the shape invariance condition. Its solutions are shape invariant superpotentials \cite{Cooper}. One solution is
\be
\label{w}
W= -i\frac{\partial}{\partial \phi} \tanh x + B{\rm sech}x~~.
\ee
In this case, an explicit calculation yields $f(-i\frac{\partial}{\partial \phi}+ c) = -2\,(-i\frac{\partial}{\partial
 \phi} )+1~.$ This implies
that we are dealing with a deformed Lie algebra with $f(J_3)=-2J_3+2c+1~.$ For
 the choice $c=-1/2$ this
is the $so(2,1)$ algebra and its realization is:
$$
J_+ =e^{i \phi} \Blb  -\frac{\partial}{\partial x} -i\frac{\partial}{\partial
 \phi} \tanh x + B{\rm sech}x  \Brb,
J_-=\! \Blb  \frac{\partial}{\partial x} -i\frac{\partial}{\partial \phi} \tanh x
 + B{\rm sech}x  \Brb e^{-i \phi},
J_3= -i\frac{\partial}{\partial \phi} +\hf.
$$
There are several other solutions
possible \cite{Gangopadhyaya_pra} and they can be derived analytically using
a point canonical transformation described in Ref. \cite{Gangopadhyaya_pct}.

The above realizations have interesting applications. The operator $J_+J_-$, 
is given by :
$$
J_+J_- = \Blb - \frac{d^2}{dx^2} + W^2  \lb x, -i \frac{\partial}{\partial \phi}-1 \rb
- \frac{dW\lb x, -i \frac{\partial}{\partial \phi}-1\rb }{dx}\Brb ~~.  
$$
When acting on factorized basis functions $e^{im\phi} \psi(x)$, one gets
$$
J_+J_- = \Blb - \frac{d^2}{dx^2} + W^2  \lb x, m-1 \rb - \frac{dW \lb x, m-1\rb }{dx}  \Brb ~,
$$
which is recognized to be the standard Hamiltonian of supersymmetric quatum mechanics.
For the choice of eq. (\ref{w}) the result is
$$
J_+J_- = \Blb - \frac{d^2}{dx^2} + (m-1)^2 \( B^2 - (m-1)^2-(m-1) \){\rm sech}^2x + B(2(m-1)+1){\rm sech}x~{\tanh}x \Brb ~~,
$$
which is just the Hamiltonian for the Scarf 
potential\footnote{\noindent
The Scarf Hamiltonian is described by a potential
$$ V_-(x,a_0,B)    = 
\left[ B^2-a_0(a_0+1) \right] {\rm sech}^2\,x + 
B(2 a_0+1) {\rm sech}\,x ~{\rm tanh}\,x +a_0^2~~.
$$
The eigenvalues of this system are given by \cite{Cooper} 
$$E_n=a_0^2-\left( a_0-n\right)^2~~.$$}
 with $m-1$ being one of the parameters. The  Scarf  potential is well known to be shape invariant, hence exactly solvable \cite{Gen}. We can also determine these eigenvalues using familiar algebraic methods of $so(2,1)$.  
The Casimir is $C \equiv J^2$ and eq. (\ref{C}) gives $J_+J_- = J_3^2-J_3-J^2$. Since the eigenvalues of $J^2, J_3$ are $j(j+1), m-1/2$ respectively, we find
$$
E=\lb m-\hf \rb ^2-\lb m-\hf \rb -j(j+1)~~.
$$  
Now substituting $j=n-m+\hf $ \cite{Adams}\,, one gets
\be
E_n ~=~ (m-1)^2 -(m-n-1)^2~~,~~n=0,1,2,\ldots
\ee
(Note that $E_0 = 0$ as expected from unbroken supersymmetric quantum mechanics.)

With a change of variable and appropriate similarity
transformations of $F(N)$ and $G(N)$\cite{Gangopadhyaya_pct}, we can relate
all solvable potentials of Ref. \cite{Cooper} to $J_+J_-$ of this
algebra and hence derive information about their spectrum algebraically.

In this paper, differential realizations in coordinate space for nonlinearly
deformed Lie algebras with three generators were obtained using bosonic
creation and annihilation operators. We have presented a unified formalism
that contains as special cases all previously given coordinate realizations
of $so(2,1),so(3)$ and their deformations. Although we have focused on deformations of the type 
specified by equation (\ref{1}), coordinate realizations for other types of deformations have also been
recently studied \cite{Angel}.


A.G. and R.D. would also like to thank the Physics Department of the University of
Illinois at Chicago for warm hospitality. Partial financial support from the U.S. 
Department of Energy and the Department of Science and Technology, Govt. of India 
(Grant No. SP/S2/K-27/94) is gratefully acknowledged.

\sss

\noindent {\bf Appendix A. Differential Realizations of $a$ and $\ad$}

In this Appendix, we discuss differential coordinate realizations of
operators $a$ and $\ad$ which satisfy the Heisenberg commutation relation
$\com{a}{\ad}=1$. The simplest choice is
\be
\label{A1} a = \dx~~,~~\ad = x~~.
\ee
As we shall see shortly, these
operators are the basic building blocks for all other realizations,
including those with higher order derivatives. Note that although the
notation $a$ and $\ad$ is being used, we are not requiring the two operators
to be Hermitian conjugates of each other.

Given any two operators $a(x, \dx)$ and $\ad(x, \dx)$ such that
$\com{a}{\ad}=1$, several simple transformations can be used to generate new
operators $\tilde a$ and $\tilde \ad$  which satisfy $\com{\tilde a}{\tilde
\ad}=1$. These transformations are:\\
(i) Rotations in the $( a, \ad) $ plane: 
$$
\tilde a = a \cos \theta  + \ad \sin \theta ~~,~~\tilde \ad = - a \sin \theta  + \ad \cos \theta ~;
$$
(ii) Change of variables $x = h(y)$: $$\tilde a(y,\dy) =
a(h(y),\frac{1}{h'(y)} \dy) ~~,~~\tilde \ad (y,\dy)=
\ad(h(y),\frac{1}{h'(y)} \dy)~, $$ where prime denotes the derivative with
respect to $y$ ;\\
(iii) Similarity transformations: 
$$
\tilde a = \phi^{-1}(x) a \phi(x)~~,~~\tilde \ad = \phi^{-1}(x) \ad \phi(x)~;
$$
(iv) Additions of arbitrary functions of the other operator:$$\tilde a = a +
\lambda(\ad)~,~ \tilde \ad = \ad~;~ \tilde a = a~,~\tilde \ad =
\ad+\mu(a)~.~$$

Successive use of the first three transformations applied to eq. (\ref{A1})
 yield
\be \label{A2}
a = \frac{\cos \theta}{h'(y)} \dy + \( h(y) \sin \theta + r(y) \cos \theta \)~~,
\ad = \frac{-\sin \theta}{h'(y)} \dy + \( h(y) \cos \theta - r(y) \sin \theta
 \)~~,
\ee
where $h(y)$ and $r(y)$ are arbitrary analytic functions of coordinate $y$.
It is easy to check that these are the most general operators linear in
$\dy$ which satisfy $\com{a}{\ad}=1$.

A natural question to ask is whether one can construct differential
coordinate realizations with second and higher order derivatives. This is in
fact possible by starting with any first order realization [say eq.
(\ref{A1}) or eq. (\ref{A2})] and using transformation (iv) to generate
higher order derivatives. For example, using eq. (\ref{A1}) and taking
$\mu(a) = a^2$ in transformation (iv) gives the realization
$$
\tilde a =\dx~~,~~\tilde \ad = x + \frac{d^2}{dx^2}~~.
$$
Although this procedure can be readily extended to get realizations of the
Heisenberg algebra involving derivatives of any desired order, it must be
kept in mind that only the realizations involving first order derivatives
are fundamental.


\end{document}